\documentstyle[aps]{revtex}

\input epsf

\begin{document}

\title{Symmetric Diblock Copolymers in Thin Films (I):\\
Phase stability in Self-Consistent Field Calculations and Monte Carlo Simulations} 

\author{
T. Geisinger, M.\ M\"{u}ller, and K.\ Binder
\\
{\small Institut f{\"u}r Physik, WA 331, Johannes Gutenberg Universit{\"a}t}
\\
{\small D-55099 Mainz, Germany}
}
\date{\today, draft to be submitted to J.Chem.Phys.}
\maketitle

\begin{abstract}

We investigate the phase behavior of symmetric $AB$ diblock copolymers confined into a thin film. The film boundaries 
are parallel, impenetrable and attract the $A$ component of the diblock copolymer. Using a self-consistent 
field technique [M.W. Matsen, J.Chem.Phys.\ {\bf 106}, 7781 (1997)], we study the ordered phases as a function 
of incompatibility $\chi$ and film thickness in the framework of the Gaussian chain model. For large film thickness 
and small incompatibility, we find first order transitions between phases with different number of lamellae which are 
parallel oriented to the film boundaries. At high incompatibility or small film thickness, transitions between 
parallel oriented and perpendicular oriented lamellae occur. 
We compare the self-consistent field calculations to Monte Carlo simulations of the bond fluctuation model
for chain length $N=32$. In the simulations we quench several systems from $\chi N=0$ to $\chi N=30$
and monitor the morphology into which the diblock copolymers assemble. Three film thicknesses are investigated, 
corresponding to parallel oriented lamellae with 2 and 4 interfaces and a perpendicular oriented morphology.
Good agreement between self-consistent field calculations and Monte Carlo simulations is found.

\end{abstract}

\section{ Introduction. }
Amphiphilic polymers are model systems for investigating mechanisms of self-assembly.
Joining chemically distinct polymers -- $A$ and $B$ -- at their ends to form an $AB$ diblock copolymer prevents macrophase separation
of the two species. In order to reduce the number of energetically unfavorable interactions between distinct blocks in a melt, the
molecules self-assemble into complex morphologies. The morphology is selected via a delicate balance between the
free energy cost of the internal interfaces and the conformational entropy loss as the molecules stretch to
fill space at constant density. The phase diagram in the bulk has been investigated in much detail as a function of the relative length 
of the blocks $f$ and the incompatibility $\chi N$\cite{Leibler,Semenov,Matsen}.  The morphologies found in copolymer melts and copolymer/homopolymer mixtures\cite{TODAY}
resemble the spatially structured phases of other amphiphilic systems (e.g., lipid/water mixtures). 

From a theoretical point of view, polymeric systems are particularly convenient for investigating mechanisms of self-assembly.
Only a small number of parameters describe the system, i.e., the fraction $f$ of $A$ monomers in the diblock, the 
molecule's end-to-end distance $R_e$ and the incompatibility $\chi N$, where $\chi$ denotes the repulsion between monomers of 
different species and $N$ the number of monomers per molecule. In general, polymeric systems are well describable by self-consistent 
field theories using the Gaussian chain model\cite{SCF1,SCF2,SCF3}. For a wide range of temperature the theory accurately calculates the excess quantities 
of the internal interfaces (e.g., the interfacial tension, the bending moduli, or the enrichment of solvent). The understanding of these interfacial
properties makes polymers suitable microscopic model systems for investigating the statistical mechanics of interacting interfaces.

If the copolymers are confined into a thin film, the interactions with the boundaries influence the morphology and its orientation.
Controlling the orientation of the spatial structure over rather large length scales is important for many practical applications\cite{Russell}.
Consequentially, much experimental effort has been directed towards a tailoring of surface properties and studying the morphologies in confined 
geometry. Lambooy {\em et al.}\cite{Lambooy} studied symmetric dPS-PMMA diblock copolymers between silicon wavers and found transitions between 
parallel oriented lamellar phases with different numbers of internal interfaces upon increasing the film thickness. Kellogg and co-workers\cite{Kellogg}
studied a similar system between surfaces, which were coated with random copolymers. This boundary mimics neutral walls and for film 
thickness corresponding to 2.5 lamellar spacings in the bulk, indeed, perpendicular oriented lamellae could be observed. The perpendicular oriented
morphology has attracted abiding interest as a template for lateral structure on the nanometer scale.

The self-assembly was analyzed for strong segregation by Walton {\em et al.}\cite{Walton} and in the framework of the self-consistent field theory by Pickett 
and Balazs\cite{Pickett} and by Matsen\cite{MSCF,REVIEW1}. Strictly neutral
walls give rise to a perpendicular orientation of the lamellae. However, surfaces that interact favorably with one component
stabilize parallel morphologies if the film thickness is compatible with the lamellar spacing in the confined state. Increasing 
the film thickness the self-consistent field studies revealed a sequence of perpendicular and parallel oriented lamellae.

The orientation of the morphologies upon confinement makes thin films a promising candidate for investigating the details of their structure via
Monte Carlo simulations. Copolymers in confined geometry have been first studied in Monte Carlo simulations by Kikuchi and Binder\cite{Kikuchi}. They found
pronounced effects of the confinement on the ordering. However, similar to experiments, Monte Carlo simulations are plagued by very long relaxation 
times and reaching thermal equilibrium is rather difficult. Indeed, the mixed state of perpendicular and parallel orientated lamellae observed 
in the simulations and experiments\cite{Koneripalli} was found to be unstable in framework of self-consistent field calculations\cite{MSCF}.
Reviews of both experiments and theory on the self-assembly of block copolymers in thin films\cite{REVIEW1,REVIEW2} give more details on previous work.

The aim of the present work is twofold: On the one hand, we calculate the phase diagram of diblock copolymers in a thin film as a function of the film 
thickness and incompatibility. We use a self-consistent field technique developed by Matsen\cite{MSCF}. Both confining walls attract the $A$ component of the diblock and 
we restrict ourselves to symmetric diblock copolymers $f=1/2$. The temperature spans the weak and intermediate segregation limit. 
On the other hand, we compare the stability of different phases in the self-consistent field calculations and in the Monte Carlo simulations
of the bond fluctuation model\cite{BFM} for chain length $N=32$ and at incompatibility $\chi N=30$. 

Our paper is arranged as follows: First we briefly describe the self-consistent field technique for diblock copolymers in thin films and introduce the model used in
the Monte Carlo simulations. Then we discuss the regions of stability of the various oriented lamellar phases as a function of the incompatibility and the film thickness. 
We present a qualitative comparison between the self-consistent field calculations and the Monte Carlo simulations and close with a brief discussion of our findings.

\section{Model and techniques}
We consider $n$ diblock copolymers containing $N=N_A+N_B$ segments in a volume $\Delta_0 \times L \times L$. $\Delta_0$ denotes the film thickness, while $L$ is the lateral extension 
of the film. The monomer number density in the middle of the film is denoted by $\rho$. The density at the film surfaces deviates from the density in the middle and it is useful to
introduce the thickness $\Delta$ of an equivalent film with constant monomer density $\Delta \equiv nN/\rho L^2$. The individual blocks of the diblock have identical length $N_A=N_B=fN$, 
such that the diblock assembles into a lamellar phase in the bulk. $R_e$ denotes the end-to-end distance of the molecule.
There is a short range repulsion between the two monomer species which can be parameterized by the Flory-Huggins parameter $\chi$.
The two surfaces of the film are impenetrable and hard. Therefore, there is no formation of islands or hole defects at the surface.  However, the free energy of a confined film
is also an important ingredient in understanding the static and dynamics of pattern formation in thin films with free boundaries\cite{BATESPRL}.
Both walls attract the $A$ component of the diblock and repel the $B$ 
component via a short range potential. 

\subsection{Self-consistent field calculations (SCF)}
The computational technique in our self-consistent field (SCF) calculations is very similar to the work of Matsen\cite{MSCF}.  In a boundary region of width $\Delta_w$ the total monomer density 
drops to zero at both walls. In our calculations we assume the monomer density profile $\rho \Phi_0$ to take the form
\begin{equation}
\Phi_0(x) = \left\{ \begin{array}{ll}
               \frac{1-\cos\left( \frac{\pi x}{\Delta_w}\right)}{2}               & \mbox{for} \qquad 0\leq x \leq \Delta_w \\
	       1                                                                 & \mbox{for} \qquad \Delta_w \leq x \leq \Delta_0 - \Delta_w \\
               \frac{1-\cos\left( \frac{\pi (\Delta_0 - x)}{\Delta_w}\right)}{2}  & \mbox{for} \qquad \Delta_0 - \Delta_w \leq x \leq \Delta_0
             \end{array}\right. 
	     \label{eqn:dens}
\end{equation}
The width and the shape of the density profile near the wall is determined by a competition of the entropy loss of the polymers near the wall (favoring a thick boundary) and
equation of state effects, which try to restore a spatially homogeneous density. The particular choice of the density profile is employed for computational convenience. The value of $\Delta_w$ 
and the shape of the profile has little effect on the relative stability of the different phases. We choose $\Delta_w=0.15 R_e$ in accord with the previous study\cite{MSCF}. This value is close to 
the ratio of the interaction range and the end-to-end distance in the Monte Carlo simulations. A film with the same number of monomers but uniform density would have the thickness 
$\Delta=\Delta_0-\Delta_w$.

Both walls attract the $A$ component of the diblock and repel the $B$ component via a short range potential. The monomer wall interaction $H$ is modeled as\cite{C1}:
\begin{equation}
H(x) = \left\{ \begin{array}{ll}
            \frac{4 \Lambda_1 b\sqrt{N}}{\Delta_w}\left\{1+\cos\left( \frac{\pi x}{\Delta_w}\right)\right\}                & \mbox{for} \qquad 0\leq x \leq \Delta_w \\
            0                                                                                                           & \mbox{for} \qquad \Delta_w \leq x \leq \Delta_0 - \Delta_w \\
            \frac{4 \Lambda_2 b\sqrt{N}}{\Delta_w}\left\{1+\cos\left( \frac{\pi (\Delta_0 - x)}{\Delta_w}\right)\right\}   & \mbox{for} \qquad \Delta_0 - \Delta_w \leq x \leq \Delta_0
        \end{array}\right.
\end{equation}
The normalization of the surface fields $\Lambda_1$ and $\Lambda_2$, which act on the monomers close to the left and the right wall, is chosen such that
the integrated interaction energy between the wall and the monomers is independent of the width of the boundary region $\Delta_w$.

The microscopic monomer densities $\hat \Phi_A$ and $\hat \Phi_B$ can be expressed as a functional of the polymer conformations $\{ {\bf r}_\alpha(\tau)\}$:
\begin{equation}
\hat \Phi_A({\bf r}) = \frac{N}{\rho} \sum_{\alpha=0}^n \int_0^f {\rm d}\tau \; \delta\left({\bf r}-{\bf r}_\alpha(\tau)\right)
\end{equation}
where the sum runs over all $n$ diblock copolymers in the system and $0 \leq \tau \leq 1$ parameterizes the contour of the Gaussian polymer. A similar expression holds
for $\hat \Phi_B({\bf r})$.
With this definition the partition function of a melt of Gaussian diblock copolymers takes the form:
\begin{eqnarray}
{\cal Z} &\sim \int {\cal D}[{\bf r}]  {\cal P}[{\bf r}] & \;\;\;\exp\left( - \rho \int {\rm d}^3{\bf r} 
                \left\{ \chi \hat \Phi_A \hat \Phi_B - H({\bf r})(\hat \Phi_A({\bf r})-\hat \Phi_B({\bf r})) \right\}\right)  \nonumber \\
             && \;\times\;\delta\left( \Phi_0({\bf r}) - \hat \Phi_A({\bf r}) - \hat \Phi_B({\bf r}) \right)
\end{eqnarray}
The functional integral ${\cal D}$ sums over all chain conformations of the diblock copolymers and 
${\cal P}[{\bf r}] \sim \exp \left(- \frac{3}{2N b^2} \int_0^1 {\rm d}\tau \;\left(\frac{{\rm d}{\bf r}}{{\rm d}\tau}\right)^2 \right)$
denotes the statistical weight of a non interacting Gaussian polymer. $b^2=R_e^2/(N-1)$ is the statistical segment length of the polymer.
This simple model neglects the coupling between the interaction energy and the chain conformations\cite{MCONF}. Thus, it cannot reproduce the stretching of the
diblock copolymer in the disordered phase observed in simulations\cite{Fried} and experiments\cite{Maurer,Stamm1,Stamm}. Moreover, the Gaussian chain model neglects a finite stiffness of 
the polymers, and the chain extensions parallel to the walls (in the parallel orientated lamellae or disordered state) always remain unperturbed.
The Boltzmann factor in the partition function incorporates the thermal repulsion between unlike monomers and the interactions between the monomers and the walls.
The last factor represents the incompressibility of the melt in the center of the film and enforces the monomer density
to decay according to Eq.(\ref{eqn:dens}) at the walls. A finite compressibility of the polymeric fluid is neglected.

Introducing auxiliary fields $W_A$, $W_B$, $\Phi_A$, $\Phi_B$ and $\Xi$ we rewrite the partition function of the multi--chain system in terms of the partition function of a single chain
\begin{equation}
{\cal Z} \sim \int {\cal D}W_A {\cal D}W_B {\cal D}\Phi_A {\cal D}\Phi_B {\cal D}\Xi \;\;\; \exp \left( -\frac{{\cal F}[W_A,W_B,\Phi_A,\phi_B,\Xi]}{k_BT}\right)
\label{eqn:Z}
\end{equation}
The free energy functional has the form:

\begin{eqnarray}
\frac{{\cal F}[W_A,W_B,\Phi_A,\phi_B,\Xi]}{n k_BT} & \equiv & - \; \ln {\cal Q}[W_A,W_B] \nonumber \\
                                                 &&         + \; \frac{1}{V} \int {\rm d}^3{\bf r} \;\; \chi N \Phi_A({\bf r}) \Phi_B({\bf r}) \nonumber \\
						 &&         - \; \frac{1}{V} \int {\rm d}^3{\bf r} \;\;  H({\bf r}) N \left\{ \Phi_A({\bf r})-\Phi_B({\bf r})\right\} \nonumber \\
						 &&         - \; \frac{1}{V} \int {\rm d}^3{\bf r} \;\;  \left\{ W_A({\bf r}) \Phi_A({\bf r}) + W_B({\bf r}) \Phi_B({\bf r})\right\} \nonumber \\
						 &&         - \; \frac{1}{V} \int {\rm d}^3{\bf r} \;\;  \Xi({\bf r}) \left\{\Phi_0({\bf r}) -  \Phi_A({\bf r}) -  \Phi_B({\bf r})\right\} 
						 \label{eqn:F}
\end{eqnarray}
where ${\cal Q}$ denotes the single chain partition in the external fields $W_A$ and $W_B$:
\begin{equation}
{\cal Q}[W_A,W_B] = \frac{1}{V} \int {\cal D}[{\bf r}] {\cal P}[{\bf r}] \;\; \exp\left( - \int_0^f {\rm d}\tau\; W_A({\bf r}(\tau)) - \int_f^1 {\rm d} \tau \; W_B({\bf r}(\tau))  \right)
\end{equation}

The functional integration in Eq.(\ref{eqn:Z}) cannot be carried out explicitly. Therefore we employ a saddlepoint approximation, which replaces the integral by the largest value
of the integrand. This maximum occurs at values of the fields and densities determined by extremizing ${\cal F}$ with respect of each of its arguments. These values are denoted by 
lower--case letters and satisfy the self-consistent set of equations:
\begin{eqnarray}
w_A({\bf r}) &=& \chi N \phi_B - H({\bf r}) N + \xi({\bf r}) \nonumber \\
w_B({\bf r}) &=& \chi N \phi_A + H({\bf r}) N + \xi({\bf r}) \nonumber \\
\phi_A({\bf r}) &=& \frac{V}{{\cal Q}} \int {\cal D}{\cal P} \;\; \int_0^f {\rm d}\tau \;\; \delta({\bf r}-{\bf r}(\tau)) \nonumber \\
                && \hspace*{1cm} \exp \left( -\int_0^f {\rm d} \tau \; w_A({\bf r}(\tau))  -\int_f^1 {\rm d}\tau\;  w_B({\bf r}(\tau)) \right) \nonumber \\
\phi_B({\bf r}) &=& \frac{V}{{\cal Q}} \int {\cal D}{\cal P} \int_f^1 {\rm d}\tau \;\; \delta({\bf r}-{\bf r}(\tau)) \nonumber \\
                && \hspace*{1cm} \exp \left( -\int_0^f {\rm d}\tau \; w_A({\bf r}(\tau))  -\int_f^1 {\rm d}\tau \; w_B({\bf r}(\tau)) \right)
\end{eqnarray}
At this stage fluctuations around the most probable configuration are ignored. Hence, the interfaces in the self-consistent (SCF) field calculations are
ideally flat and there is no broadening by fluctuations of the local position of the interfaces (capillary waves).

The distribution of copolymer segments is calculated by solving appropriate diffusion equations for the end segment distributions.
In order to study the self-assembly into various morphologies, we expand the spatial dependence of the densities and fields in a set of orthonormal functions 
that possess the symmetry of the morphology being considered. This results in a set of non--linear equations which are solved by a Newton-Raphson like method.
Substituting the saddlepoint values of the densities and fields into the free energy functional (\ref{eqn:F}) we calculate the free energy of the various morphologies.
For the perpendicular oriented lamellar phase the value of the free energy has to be minimized with respect to the lamellar spacing.
We use up to 220 basis functions for the calculation of the perpendicular oriented lamellar phases and thus achieve a relative accuracy of the order $10^{-4}$ for the free energy.

\subsection{Monte Carlo simulations (MC)}
For the Monte Carlo (MC) simulations we employ the bond fluctuation model\cite{BFM}. This coarse grained lattice model captures the relevant universal features of polymeric materials:
excluded volume of segments, chain connectivity, and a short range thermal interaction. Many thermodynamic properties of the model have been determined in previous studies\cite{MREV}
and the model is a good compromise between the computational advantages of a lattice model and a faithful representation of continuum space properties. In particular, the 
relation between the model parameters, the local fluid--like packing of monomers, and the phase behavior has been investigated in detail.
Within the framework of this model a small number of chemical repeat units is represented by the eight corners of a cube on a three dimensional lattice. Monomers along a polymer 
are connected via one of 108 bond vectors of length $2,\sqrt{5},\sqrt{6},3$, and $\sqrt{10}$. The distances are measured in units of the lattice spacing. The polymers comprise $N=32$
monomers. We work at a monomer number density $\rho=1/16$. This value corresponds to a concentrated solution or a melt. Under these conditions the  end-to-end distance is
$R_e \approx 17$ and, in accord with previous studies, we use $b=R_e/\sqrt{N-1}=3.05$ for the statistical segment length in the SCF calculations\cite{OLD}.

One half of the polymer consists of $A$ monomers, the other consists of $B$ monomers. Monomers of the same type attract each other via a square well potential, while there is
a repulsion between unlike species. The interaction range comprises the 54 nearest neighbors up to a distance $\sqrt{6}$. This corresponds roughly to the first neighbor shell 
in the monomer density pair correlation function. The well depths of the interactions are chosen symmetrically: $\epsilon_{AA}=\epsilon_{BB}=-\epsilon_{AB} \equiv -\epsilon$.

The phase behavior and structure of a binary blend of $A$ and $B$ homopolymers\cite{MW,REV} has been investigated in the framework of this model. The phase diagram of binary
homopolymer mixtures and ternary homopolymer/copolymer blends\cite{MS1} as well as the interfacial structure between  unmixed phases are well describable by the Gaussian chain model 
if the Flory-Huggins parameter is identified via\cite{M0}
\begin{equation}
\chi = \frac{2 \epsilon}{k_BT} \int_{r \leq \sqrt{6}} {\rm d}^3{\bf r}\;\;g^{\rm inter}({\bf r}) \approx \frac{2 \epsilon}{k_BT}\;2.65 \qquad ,
\label{eqn:chi}
\end{equation}
where we have used the symmetry of the monomeric interactions. Here, $g^{\rm inter}$ denotes the intermolecular pair correlation function in the melt and the integral is extended 
over the range of the square well potential. Moreover, we have assumed that $g^{\rm inter}$ is largely independent from temperature. We use the value 2.65 for the number of 
monomers of different chains in the range of the square well interaction. This identification of the Flory-Huggins parameter is based on the energy of mixing; entropic contributions 
to the free energy due to packing effects or conformational changes are negligible. We perform Monte Carlo (MC) simulations at $\epsilon=0.1769 k_BT$. This value corresponds to $\chi N = 30$.

The simulation cell possesses a geometry of the form $\Delta_0 \times L \times L$. Periodic boundary conditions are applied in the lateral directions.
The two impenetrable walls at $x=0$ and $x=\Delta_0+1$ interact with monomers in the nearest two layers via a square well potential. An $A$
monomer in the interaction range of the walls lowers the energy by an amount $\epsilon_w$ while a $B$ monomer increases the energy by the same 
amount. If the surface is completely covered with $A$ monomers the wall interaction energy per chain is:
\begin{equation}
\frac{F_{\rm wall}}{n k_BT} = -2 \times \frac{2\epsilon_w N}{\Delta k_BT}
\label{eqn:ewall}
\end{equation}
In the simulations we use the value $\epsilon_{w}=0.1 k_BT$. Simulations\cite{WET} of binary blends (with identical interactions) show, that for this strength of surface interactions the
$A$ component wets the surface for $\epsilon<0.043 k_BT$ or $\chi N<7.3$. For the comparison between the SCF calculations and the MC simulations we adjust 
the surface fields $\Lambda$ as to result in the same contribution to the energy. Hence, the surface interactions in the MC simulations correspond to $\Lambda_1N=\Lambda_2N=0.375$
(cf.\ Eq.(\ref{eqn:FW}) below).

The MC simulations comprise three different moves: The conformations of the polymers are updated via local hopping of the monomers and slithering snake like motions.
In the former, we randomly choose a monomer and try to displace it by one lattice unit in a random direction. During the slithering snake attempts, we randomly choose a chain end
and try to attach it at the opposite end of the chain. The monomer identity ($A$ or $B$) of the chains is correspondingly updated to precisely conserve the composition of the chain\cite{Fried}.
The latter moves relax the conformations of the polymers a factor of the order $N$ faster than local updates.
Moreover, we allow for $A \rightleftharpoons B$ flips, in which the identity of the $A$ and $B$ monomers of a randomly chosen  polymer are exchanged. 
One Monte Carlo step consists of 3 slithering snake attempts per chain, 1 local hopping attempt per monomer and 1  $A \rightleftharpoons B$ flip per diblock.
Every 12500 Monte Carlo steps a configuration was stored for further analysis. 
Since we are interested in studying the stability of different morphologies, we do not impose a specific morphology on the starting configurations\cite{Grest}.
Rather, we let the structure self-assemble via a quench from the disordered phase to the ordered state at $\chi N=30$, and monitor the morphologies which
occur in several independent runs with identical parameters. 

\section{ Phase diagram. }
At high incompatibility, many features of confined diblock copolymers can be deduced from the strong stretching theory\cite{Semenov}. This has been applied to
study the effect of confinement by Turner\cite{Turner} and Walton {\em et al}\cite{Walton} (see also \cite{Kikuchi}). We follow the notation of Ref.\cite{MSCF}.
We consider a parallel lamellar phase $L_p$ with $p$ internal interfaces.
In the limit $\chi N \gg 10$ the $A$ and $B$ rich domains are well segregated and the junction points are confined to a narrow interfacial region. 
To fill space uniformly the copolymers stretch. In a lamellar morphology each half of the diblock forms a brush. These brushes do not interpenetrate.
In the parallel lamellar phase $L_p$ with $p$ interfaces each brush has the height $\Delta/2p$ and the free energy cost due to the stretching of the chains in the brush\cite{MWC} amounts to:
\begin{equation}
\frac{F_{\rm brush}}{nk_BT} = 2 \times \frac{\pi^2 (\Delta/2p)^2 }{8 (N/2) b^2}
\end{equation}
Each half of the diblock covers an interfacial area $pN/\rho\Delta$.  Estimating the value of the interfacial tension between $A$ and $B$  domains by the interfacial tension in a binary blend 
$\sigma/k_BT \approx \rho b \sqrt{\chi/6}$, the free energy contribution of the internal interfaces per polymer takes the form:
\begin{equation}
\frac{F_{\rm inter}}{nk_BT} = \sqrt{\chi/6} \frac{p b N}{\Delta}
\end{equation}
The balance between these two terms determines the behavior in the bulk. This leads to a prefered lamellar spacing $D_{\rm b} = 2\Delta_{\rm min}/p= 2(8\chi N/3\pi^4)^{1/6}R_e$.
If the walls preferentially interact with one component the interaction energy of the monomers with the walls gives another contribution to the
free energy. Using the expression for the density profiles and the wall monomer interactions, this contribution takes the form:
\begin{equation}
\frac{F_{\rm wall}}{n k_BT} = \frac{1}{N} \int {\rm d}^3{\bf r}\; H({\bf r}) \rho \Phi_0({\bf r}) = -\frac{\Lambda_1 N + \Lambda_2 N}{\Delta/(b\sqrt{N})}
\label{eqn:FW}
\end{equation}
where we have assumed that each surface is covered completely with the energetically favored component. For parallel lamellar phases with an 
odd or an even number of interfaces exposed to symmetric or antisymmetric surfaces fields, respectively, the contributions cancel.

The confinement into a film also reduced the conformational entropy of the molecules. Using ground state dominance, the entropy of an inhomogeneous melt
is given by\cite{Lifshitz}:
\begin{equation}
\frac{F_{\rm conf}}{nk_BT} =  \frac{b^2}{24 n} \int {\rm d}{\bf r} \frac{\left(\nabla \Phi_0({\bf r})\right)^2}{ \Phi_0({\bf r})} = \frac{\pi^2 N b^2}{24 \Delta_w \Delta}
\end{equation}
This contributions depends strongly on the detailed density profile at the wall. However, it does not discriminate between the different phases and, hence, is
irrelevant to the stability of the different morphologies.

In the perpendicular lamellar phase $L_\perp$ both components cover an equal amount of surface area and, hence, the contribution of the surface fields vanishes. 
Moreover, the lamellar period is free to adjust
as to minimize the contribution from the internal interfaces and the chain stretching. As a result, the lamellar spacing in the strong stretching theory is identical in the perpendicular morphology
of a film and in the bulk.

\begin{figure}[htbp]
    \begin{minipage}[t]{160mm}%
       \mbox{
        \setlength{\epsfxsize}{8cm}
        \epsffile{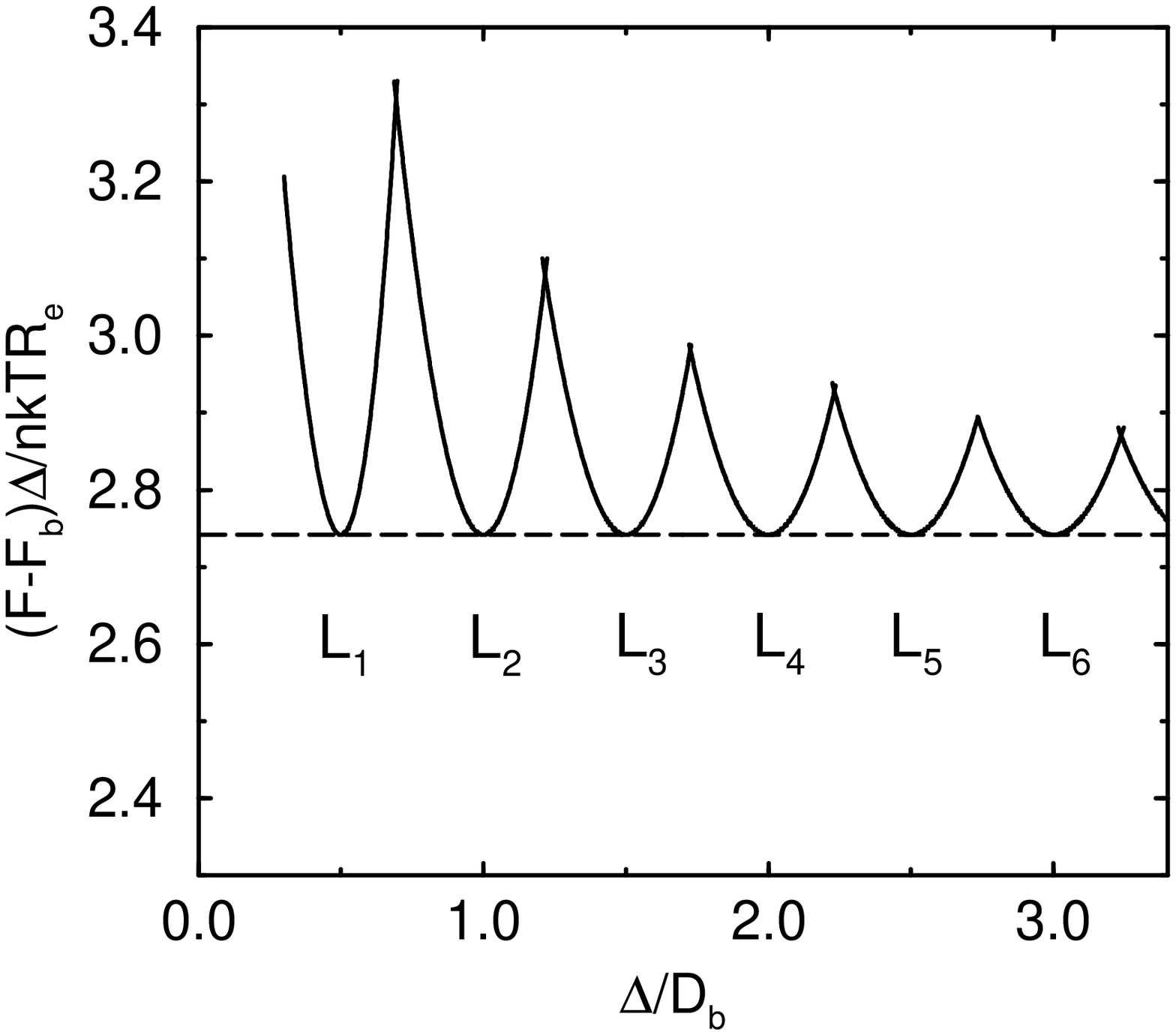}
       }\\
       \mbox{
        \setlength{\epsfxsize}{8cm}
        \epsffile{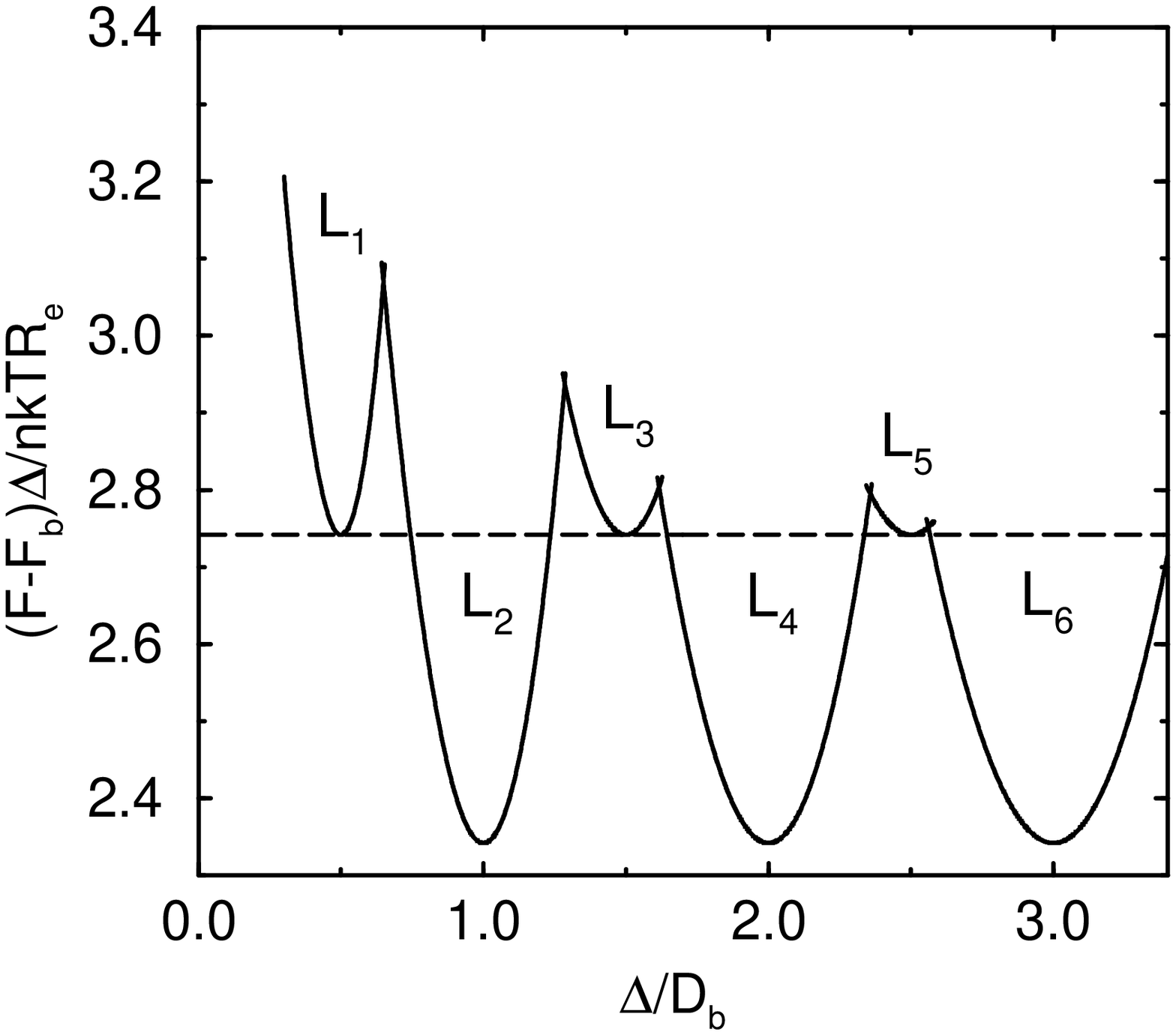}
       }\\
       \mbox{
        \setlength{\epsfxsize}{8cm}
        \epsffile{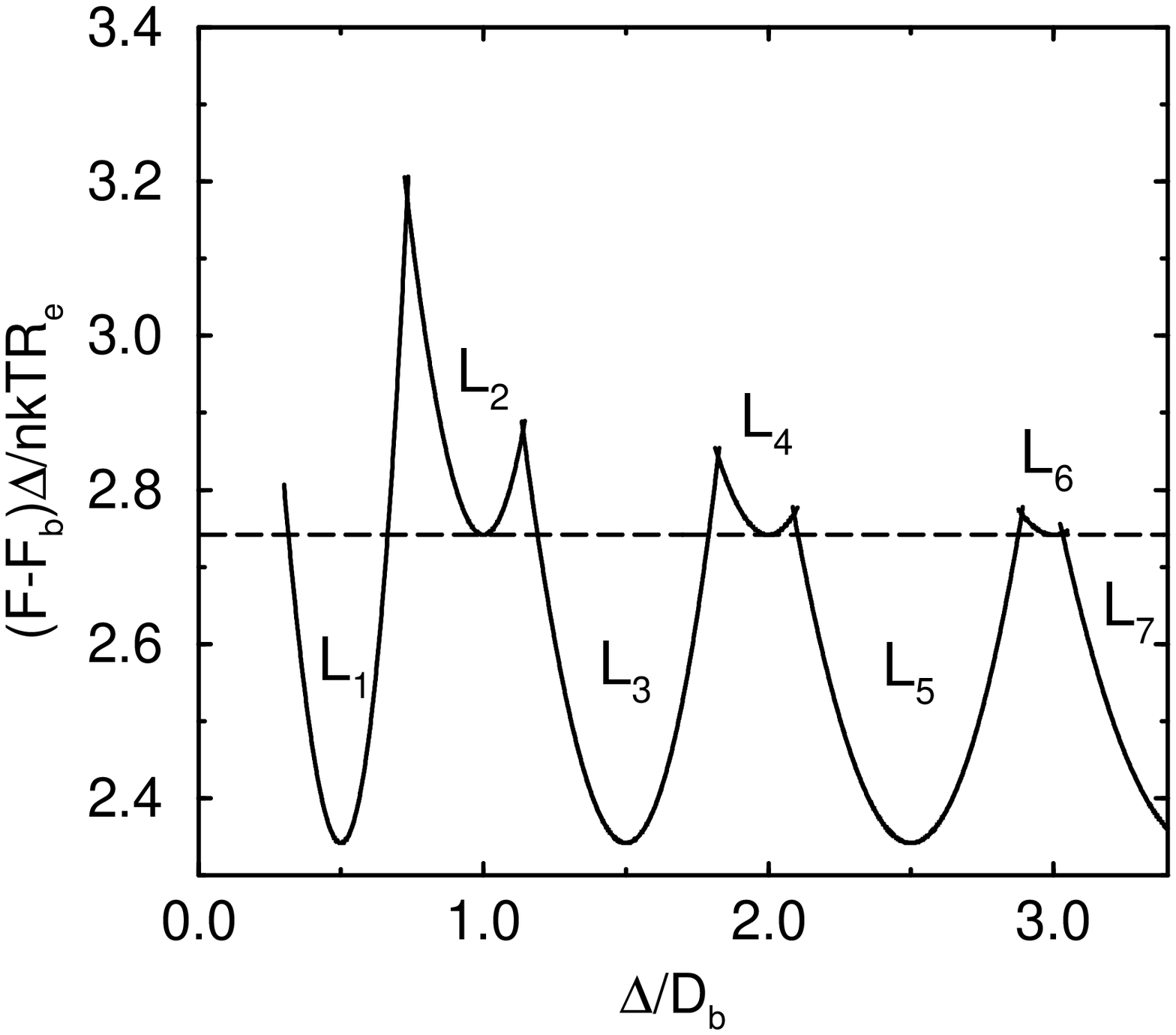}
       }
    \end{minipage}%
    \hfill%
    \begin{minipage}[b]{160mm}%
    \vspace*{1cm}
       \caption{
       \label{fig:SST} Free energy as predicted by the strong stretching theory as a function of the film thickness at $\chi N = 30 $.
       The solid lines correspond to parallel lamellar phases $L_p$, where $p$ denotes the number of $AB$ interfaces. The dashed line represents the perpendicular
       lamellar phase.
       ({\bf a}) neutral walls $\Lambda_1 N = \Lambda_2 N = 0$
       ({\bf b}) symmetric walls $\Lambda_1 N = \Lambda_2 N = 0.2$, and
       ({\bf c}) antisymmetric walls $\Lambda_1 N = -\Lambda_2 N = 0.2$
       }
    \end{minipage}%
\end{figure}

Fig.\ref{fig:SST} presents the excess free energy for the different phases in the strong stretching approximation for $\chi N=30$. 
We plot the difference $(F-F_{\rm b})\Delta/nk_BTR_e$, which is proportional to the difference between the free energy in the 
confined geometry and the bulk free energy per unit area of the film.
For neutral walls ({\bf a}), which do not prefer a component, the $L_\perp$ phase
is stable for all film thicknesses. The parallel lamellar morphology has the identical free energy only 
if the film thickness coincides with half integer multiples of the bulk period $D_{\rm b}$. For symmetric walls ({\bf b}) the free energy of
the parallel morphology with an even number of interfaces is lowered, because the prefered component is brought into contact 
with both walls. The free energies of the other phases remain unchanged.
Upon increasing the film thickness one observes transitions from morphologies with an
even number $2p$ of interfaces at film thickness around $p D_{\rm b}$ and perpendicular orientated lamellae for thickness $(2p+1)D_{\rm b}/2$.
In the case of antisymmetric surface fields ({\bf c}) the free energy of the parallel lamellae with an odd number (2p+1) of interfaces is lowered,
and one finds transitions between those parallel lamellae for film thickness close to $(2p+1)D_{\rm b}/2$ and the perpendicular morphology
for thickness around integer multiples of the bulk period.

\begin{figure}[htbp]
    \begin{minipage}[t]{160mm}%
       \setlength{\epsfxsize}{8cm}
       \mbox{
        \setlength{\epsfxsize}{8cm}
        \epsffile{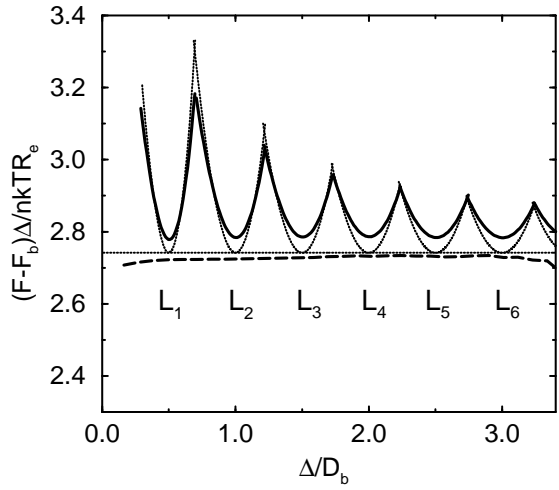}
       }\\
       \mbox{
        \setlength{\epsfxsize}{8cm}
        \epsffile{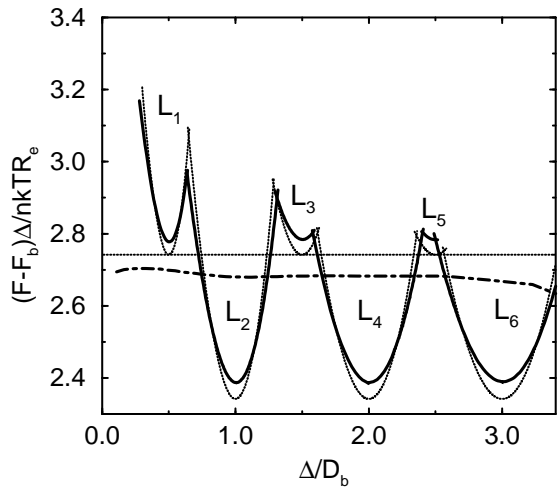}
       }
    \end{minipage}%
    \hfill%
    \begin{minipage}[b]{160mm}%
    \vspace*{1cm}
       \caption{
       \label{fig:SST_SCF} Comparison between the strong stretching theory (dotted lines) and the self-consistent field theory (thick solid lines ($L_p$) and
       dashed dotted line ($L_\perp$)) at $\chi N=30$.
       ({\bf a}) neutral walls $\Lambda_1 N = \Lambda_2 N = 0$ and
       ({\bf b}) symmetric walls $\Lambda_1 N = \Lambda_2 N = 0.2$
       }
    \end{minipage}%
\end{figure}

In Fig \ref{fig:SST_SCF} we compare the results with the full SCF calculations. In this representation the curves are
qualitatively similar. In particular, both approaches yield the same sequence of morphologies as the film thickness is increased. However, 
the absolute values of the free energies differ by more than $20\%$ for these parameters. Moreover, there are some subtle differences: For 
neutral walls ({\bf a}), the free energy of the perpendicular morphology is strictly lower than that of the parallel oriented lamellae. 
This is in accord with the calculations of Pickett and Balazs\cite{Pickett}. 
For symmetric surface fields ({\bf b}) the free energy of the parallel morphology with an even number of interfaces is lowered. 
The value of the relative shifts upon increasing the surface field in the strong stretching theory and the SCF calculations agree 
nicely. In both approaches the free energy of the parallel morphology with an odd number of lamellae remains almost unaffected. 
This indicates that the structure at the surfaces is hardly perturbed by the weak surface fields.
However, the free energy
of the perpendicular morphology is independent of the surface fields in the strong stretching theory, while the presence of the surface fields lowers
the free energy in the SCF calculations, indicating a dependence of the spatial arrangement on the surface fields.

\begin{figure}[htbp]
    \begin{minipage}[t]{160mm}%

       \mbox{
        \setlength{\epsfxsize}{8cm}
        \epsffile{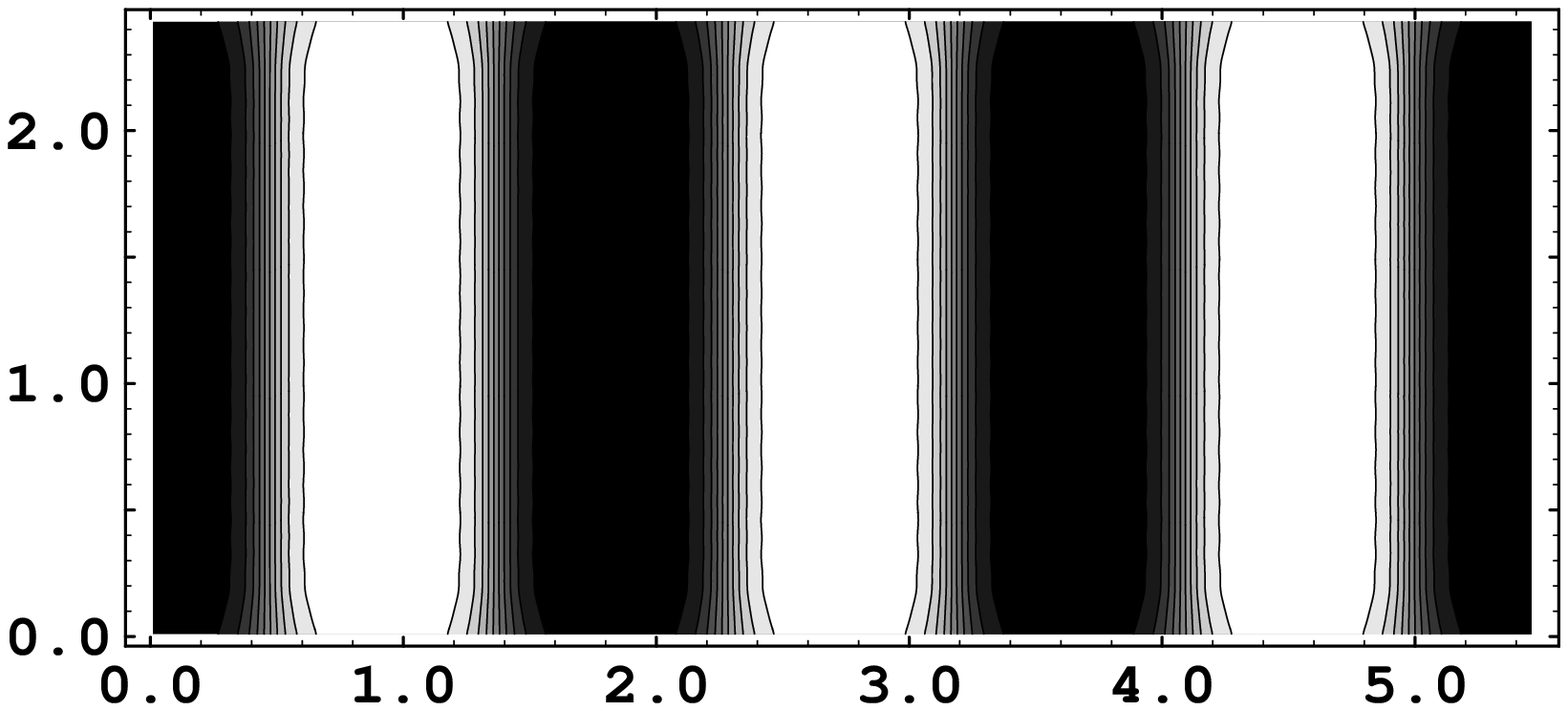}
       }

       \vspace*{1cm}

       \mbox{
        \setlength{\epsfxsize}{8cm}
        \epsffile{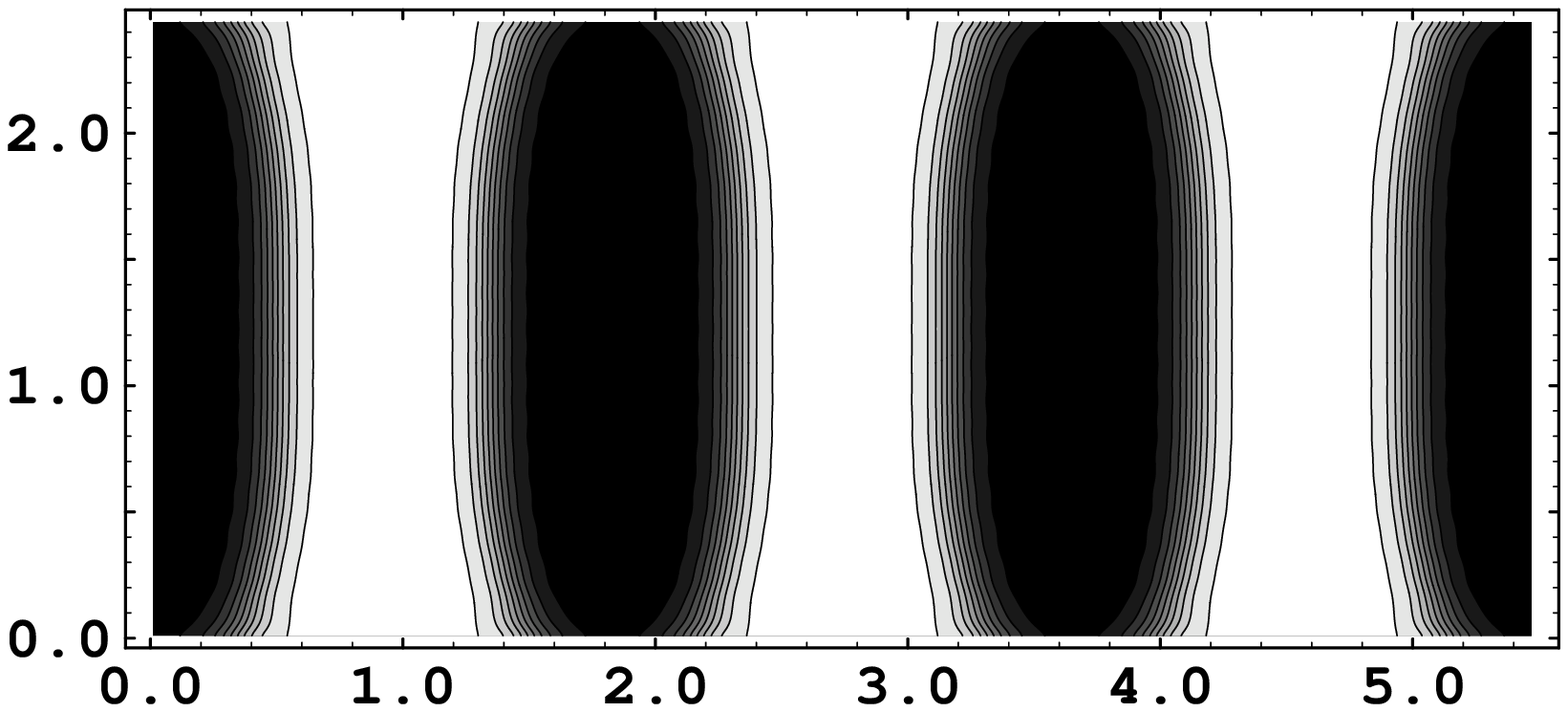}
       }
    \end{minipage}%
    \hfill%
    \begin{minipage}[b]{160mm}%
    \vspace*{1cm}
       \caption{
       \label{fig:SCF_CONT} Contour plots of the composition in the perpendicular lamellar phase at $\chi N=30$ and $\Delta/R_e=2.3$.
       $A$ rich regions are light, $B$ rich regions are dark.
       ({\bf a}) neutral walls $\Lambda_1 N = \Lambda_2 N = 0$ and
       ({\bf b}) symmetric walls $\Lambda_1 N = \Lambda_2 N = 0.2$
       Lines correspond to constant compositions in the $xz$ plane. The coordinate $x$ across the film is oriented along the ordinate here.
       (Lengths are measured in units of $R_e$.)
       }
    \end{minipage}%
\end{figure}

The composition profiles of the perpendicular morphology are presented in Fig.\ref{fig:SCF_CONT} for neutral walls ({\bf a}) and symmetric walls attracting the 
$A$ component ({\bf b}).  The $A$ rich regions are bright and $B$ rich regions are shaded darkly. In the case of neutral walls, the interface between the $A$ and $B$ domains
runs strictly perpendicular to the surface. However, the interfacial width broadens close to the surface. Partially, this is due to the reduction of the density 
in the surface region, which reduces the incompatibility between the two components. This reduces the $AB$ interfacial tension as the interface intersects the wall.
Moreover, the polymers in the vicinity of the surfaces are aligned parallel to the wall, and this orientation is compatible with their conformation at the $AB$ interface 
as it approaches the wall\cite{Pickett}. Both effects reduce the free energy of the $AB$ interface in the perpendicular morphology. This gives rise to a 
negative line tension\cite{MSCF} and tends to stabilize the perpendicular phase.

Upon increasing the surface interactions ({\bf b}), the $AB$ interface bends and intersects the wall at an angle. This distortion of the interface close to the surface 
increases the surface area covered by the energetically favored $A$ component and lowers the free energy of the perpendicular phase\cite{Pickett}. This effect is not captured by the strong stretching 
approximation. However, Pereira and Williams\cite{Williams} have argued that it remains typically small. Note that the surface fields are rather small such that the $A$ 
component does not wet the surface.

\begin{figure}[htbp]
    \begin{minipage}[t]{160mm}%
       \setlength{\epsfxsize}{10cm}
       \mbox{
        \epsffile{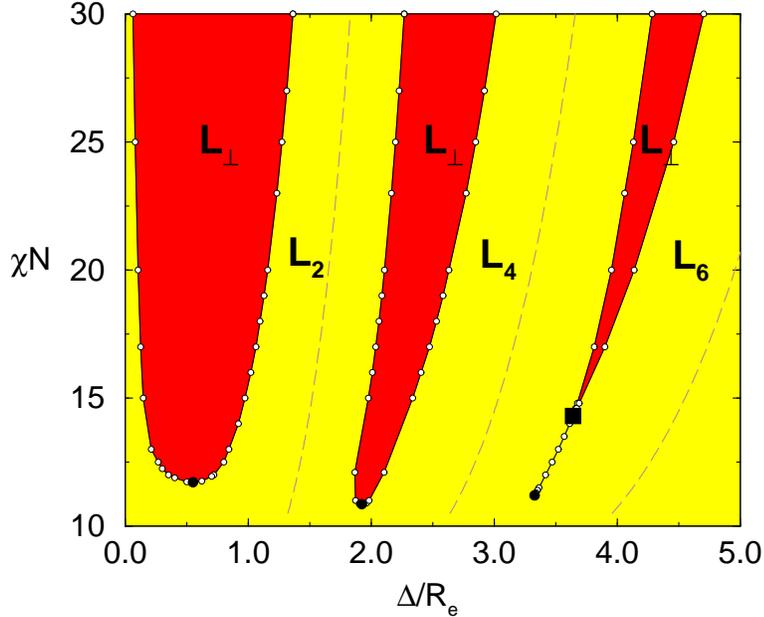}
       }
    \end{minipage}%
    \hfill%
    \begin{minipage}[b]{160mm}%
    \vspace*{1cm}
       \caption{
       \label{fig:PD} Phase diagram of a thin film with symmetric walls $\Lambda_1 N = \Lambda_2 N = 0.2$ as a function of the 
       incompatibility $\chi N$ and the film thickness $\Delta/R_e$.  $L_2$, $L_4$, and $L_6$ denote parallel lamellar phases
       with $2$,$4$, and $6$ $AB$ interfaces, whereas $L_\perp$ denotes the perpendicular lamellar phase. The dashed lines mark
       multiples of the bulk lamellar period. The square denotes the approximate location of the triple point.
       }
    \end{minipage}%
\end{figure}

The phase diagram as a function of the incompatibility $\chi N$ and film thickness is presented in Fig.\ref{fig:PD} for symmetric boundary fields. At high incompatibilities we find a sequence of
perpendicular aligned lamellae $L_\perp$ and parallel lamellae $L_p$ with an even number of interfaces. The latter are stable for film thicknesses close to integer multiples of
the bulk period. Upon increasing the film thickness the stability region of the perpendicular oriented morphology decreases. The free energy difference 
between the two morphologies is related to the balance between the surface interactions which favor parallel orientation and the free energy costs of imposing a lamellar spacing
which differs from the prefered bulk value.
The surface contribution is independent of the film thickness. The free energy costs per lamellae due to a mismatch in the lamellar spacing increase
quadratically.  Hence, it is favorable to distribute the mismatch evenly among the p lamellae; the mismatch per lamellae is proportional to $1/p$. Therefore the 
free energy of the film due to deviations of the film thickness from the prefered spacing decreases like $p \times (1/p)^2$; i.e., for thick films
the mismatch becomes unimportant and only parallel lamellae are stable.

In thin films the translational symmetry parallel to the walls is spontaneously broken and the SCF theory predicts a second order
transition from the disordered state to the perpendicular lamellar phase. 
For thicker films we find direct transitions between the parallel lamellar phase with 4 and 6 interfaces. Only at higher incompatibilities we encounter a triple point,
at which the two parallel lamellar phases $L_4$ and $L_6$ coexist with a  perpendicular aligned lamellar phase $L_\perp$. At higher incompatibilities we find the
sequence $L_4$,$L_\perp$ and $L_6$ upon increasing the film thickness. We expect this behavior to be representative for larger film thicknesses and the incompatibility
at which the triple point occurs increases with the film thickness.
An important point is that the theory predicts a gradual onset of parallel ordering ($L_p$) as $\chi N$ increases, without a phase transition from the disordered phase. This happens 
because for finite $\Lambda$ the surface fields create surface induced order of lamellar type already in the disordered phase, and this order gets gradually stronger as $\chi N$ increases.

\begin{figure}[htbp]
    \begin{minipage}[t]{160mm}%


       \vspace*{1cm}

       \mbox{
        \setlength{\epsfxsize}{8cm}
        \epsffile{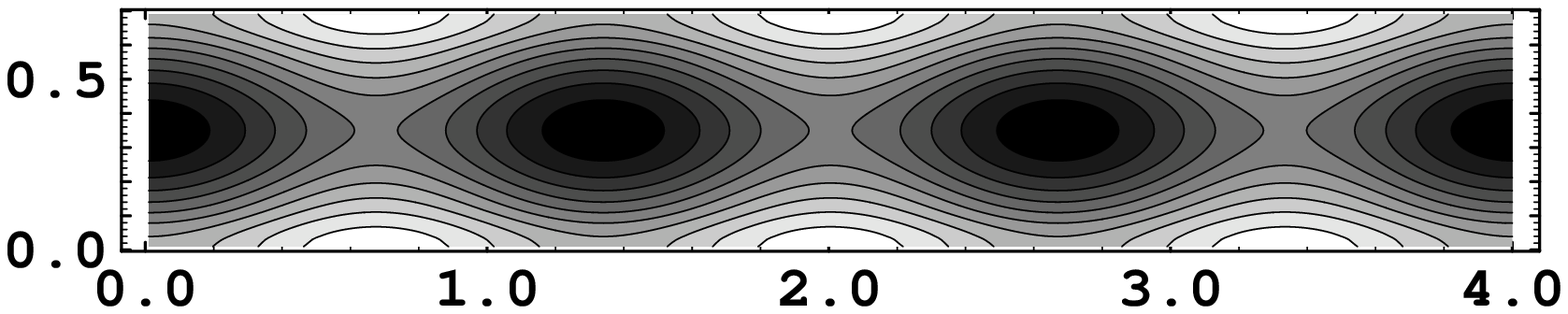}
       }

       \vspace*{1cm}

       \mbox{
        \setlength{\epsfxsize}{8cm}
        \epsffile{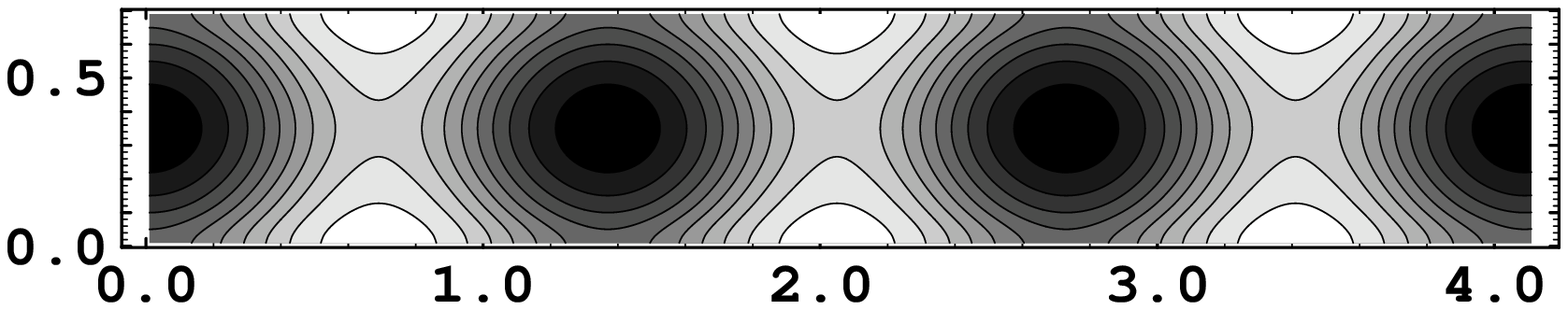}
       }
    \end{minipage}%
    \hfill%
    \begin{minipage}[b]{160mm}%
    \vspace*{1cm}
       \caption{
       \label{fig:finger1} Contour plots of the composition in a symmetric film $\Delta/R_e=0.55$ and $\Lambda_1 N = \Lambda_2 N = 0.2$.
       Upon increasing the incompatibility, one encounters a second order transition to a perpendicular lamellar phase. The critical point occurs at about
       $\chi_c N =  11.72$.
       ({\bf a}) $\chi N=11.72$,  (Figure available upon request)
       ({\bf b}) $\chi N=11.8$, and
       ({\bf c}) $\chi N=12.5$
       }
    \end{minipage}%
\end{figure}

\begin{figure}[htbp]
    \begin{minipage}[t]{160mm}%

       \mbox{
        \setlength{\epsfxsize}{8cm}
        \epsffile{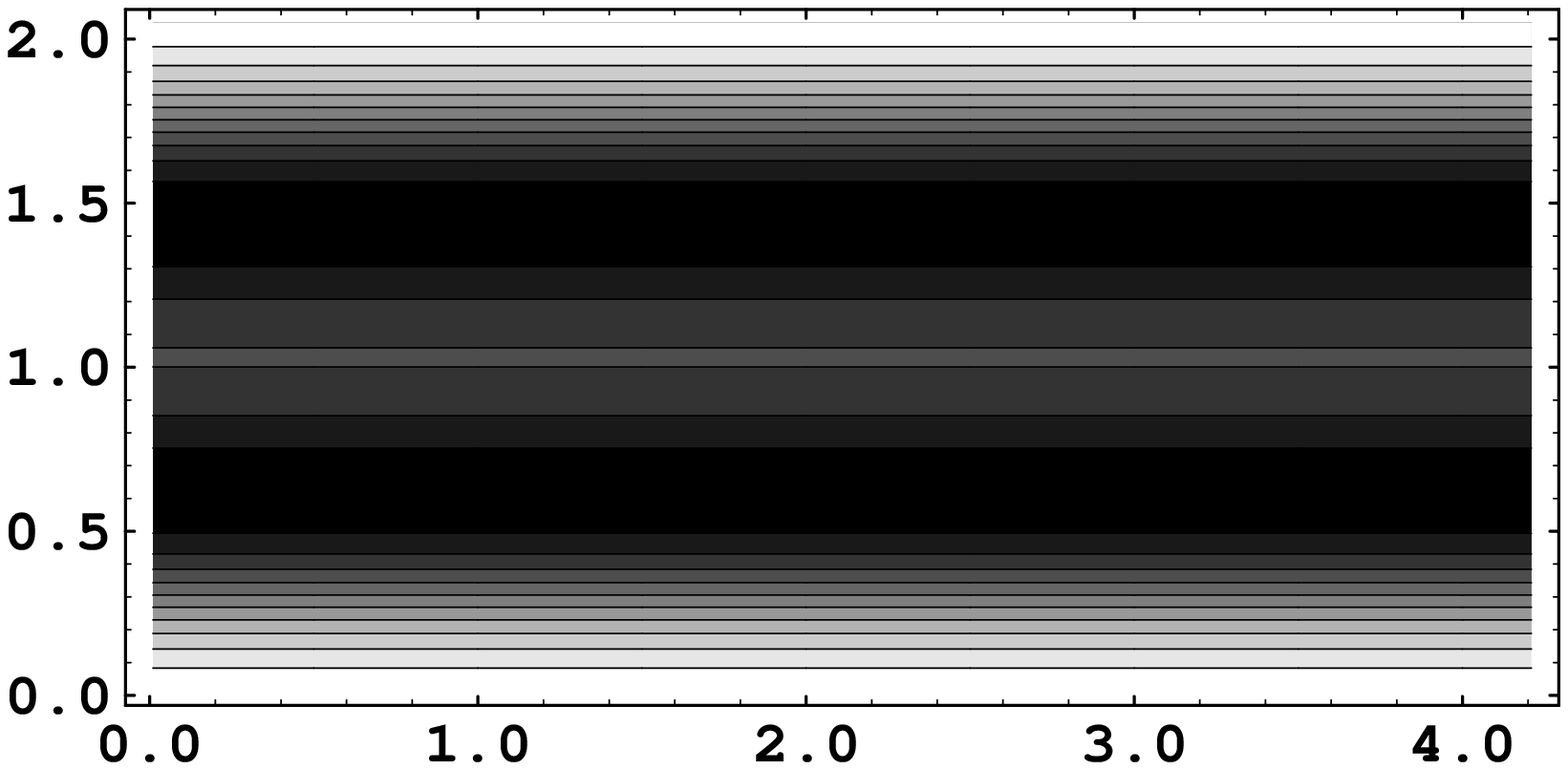}
       }

       \vspace*{1cm}

       \mbox{
        \setlength{\epsfxsize}{8cm}
        \epsffile{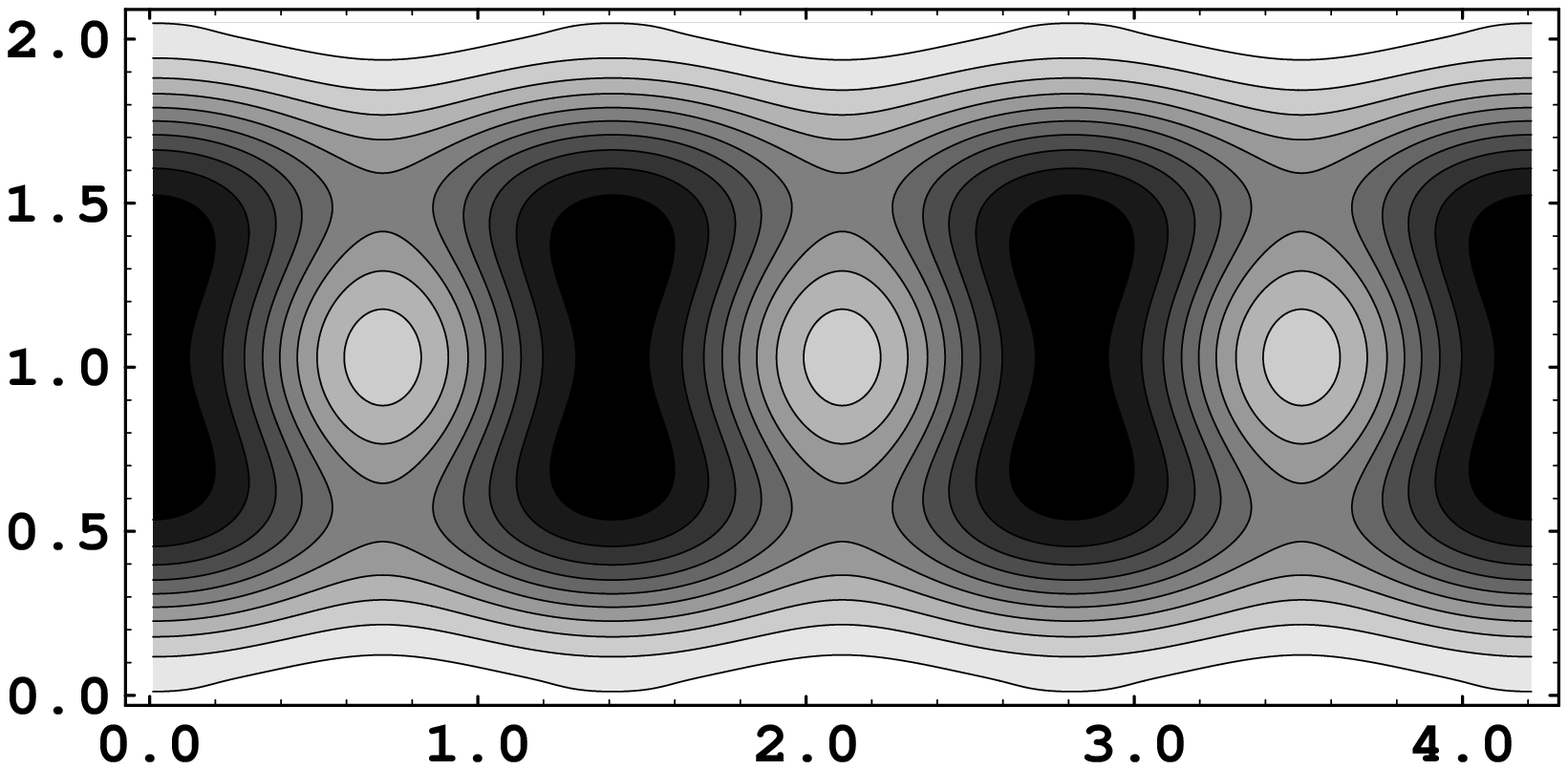}
       }

       \vspace*{1cm}

       \mbox{
        \setlength{\epsfxsize}{8cm}
        \epsffile{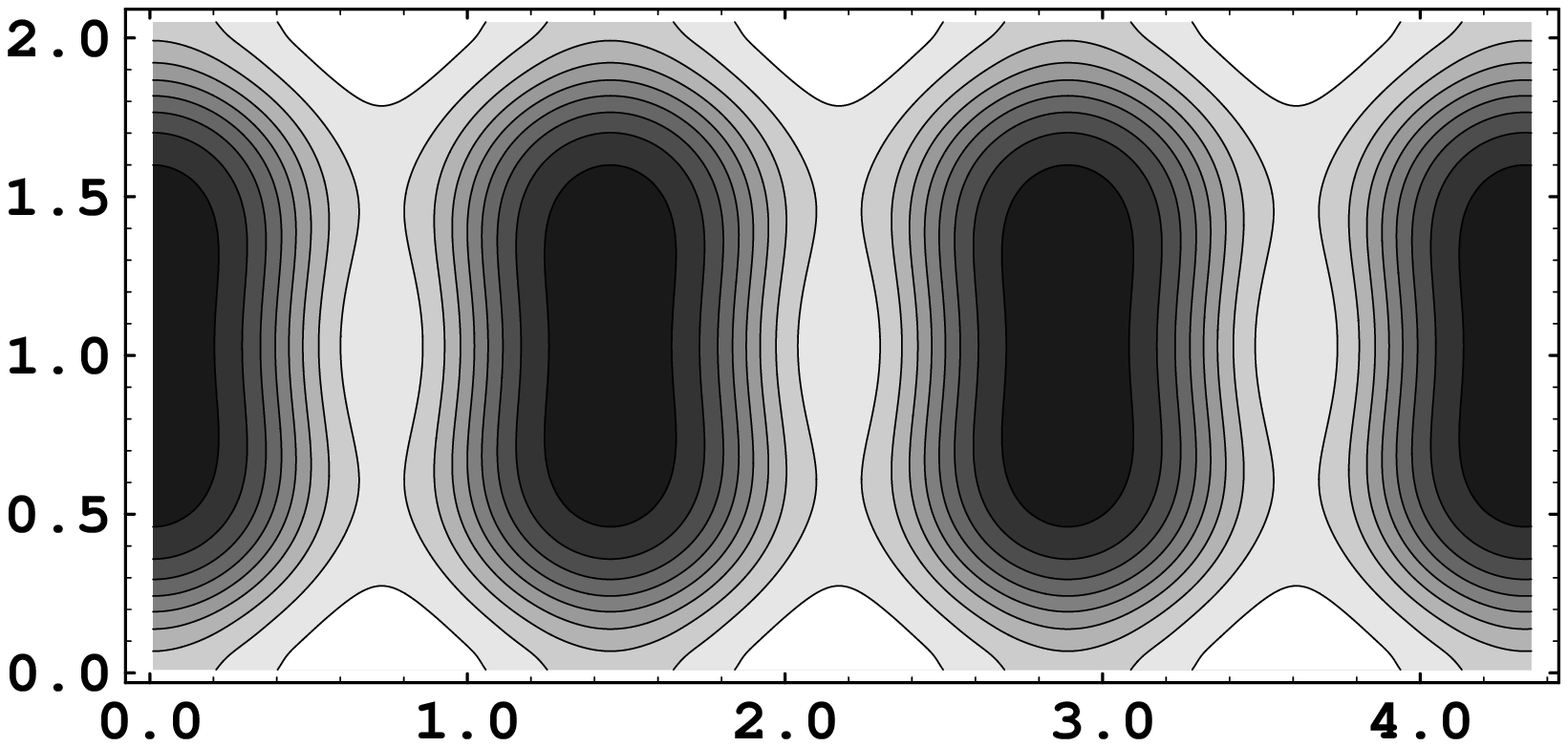}
       }
    \end{minipage}%
    \hfill%
    \begin{minipage}[b]{160mm}%
    \vspace*{1cm}
       \caption{
       \label{fig:finger2} Contour plots of the composition in a symmetric film $\Delta/R_e=1.92$ and $\Lambda_1 N = \Lambda_2 N = 0.2$.
       Upon increasing the incompatibility, one encounters a second order transition to a perpendicular lamellar phase.
       The critical point occurs at about  $\chi_c N =  10.87$.
       ({\bf a}) $\chi N=10.85$,
       ({\bf b}) $\chi N=11.5$, and
       ({\bf c}) $\chi N=13$
       }
    \end{minipage}%
\end{figure}

The composition profiles in a thin film close to these critical points, where the perpendicular lamellae emerge, are presented in Fig.\ref{fig:finger1} 
for a film thickness $\Delta/R_e=0.55$ and Fig.\ref{fig:finger2} for $\Delta/R_e=1.92$. When the incompatibility is increased the perpendicular modulations become
more pronounced. Slightly above the critical points the $B$--rich domains for $\Delta/R_e=0.55$ and the $A$--rich domains for $\Delta/R_e=1.92$ form cylinders
which run parallel to the surfaces. This behavior resembles the fingerprint--like morphology observed in experiments of Chaikin and co-workers\cite{Chaikin}. For slightly 
asymmetric diblocks or stronger surface fields even more pronounced effects could be anticipated. However, the neglect of fluctuations imparts a quantitative 
inaccuracy to the SCF calculations in the weak segregation limit. In particular, the existence of critical points where a second order transition from the disordered phase into 
the perpendicular oriented lamellar structure occurs is questionable. In the bulk, one encounters a fluctuation--induced first order transition rather than a critical point\cite{Fredrickson}.

\section{Comparison between self-consistent field (SCF) calculations and Monte Carlo (MC) simulations. } 
\begin{figure}[htbp]
\begin{minipage}[t]{160mm}%
   \mbox{
   \setlength{\epsfxsize}{8cm}
   \epsffile{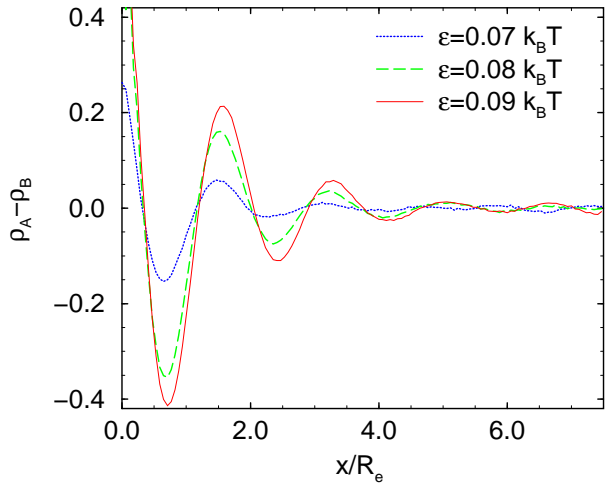}
   }
   \end{minipage}%
  \hfill%
   \begin{minipage}[b]{160mm}%
   \caption{
   \label{fig:chic} 
   Difference between the $A$ and $B$ monomer density in the vicinity of the surface. Upon approaching the
   order-disorder transition, we observe that the correlation length and amplitude of composition oscillation increase.
   }
\end{minipage}%
\end{figure}

In the following we compare our SCF calculations at $\chi N=30$ and $\Lambda_1N=\Lambda_2 N=0.375$ to the corresponding MC simulations in the framework of the
bond fluctuation model. The value of the incompatibility lies in the regime of intermediate segregation and is well inside the experimentally accessible range. For much smaller incompatibilities
there are strong fluctuation effects. We have performed some preliminary simulations in a $256 \times 64 \times 64$ geometry with $\epsilon_w = 0.1 k_BT$ to study the ordering behavior.
The results for the difference between the $A$ and $B$ monomer density in the vicinity of the surfaces are displayed in Fig.\ref{fig:chic}. Upon increasing the incompatibility,
the amplitude and correlation length of composition fluctuations increase. Moreover, the period of the oscillations increases slightly upon increasing the incompatibility. This
indicates a stretching of the molecules. For $\epsilon=0.09$ we observe a weak modulation of the composition across the whole film thickness.
Hence, we expect the order-disorder transition to occur in the range $\epsilon \approx 0.09(1)$ or $13.5 \leq \chi N \leq 17$\cite{C2}. This is in accord with a previous estimate of the transition 
temperature\cite{MS1}. Moreover, the magnitude of this shift in the transition temperature is in qualitative agreement with fluctuation corrections calculated by Fredrickson and Helfand\cite{Fredrickson}. 
They predict $\chi_c N = 10.5+41/(R_e^2\rho^{2/3})^{1/3}$ for very long chains. Deviations of similar magnitude, though by a different mechanism, have been predicted in the framework of the P-RISM
theory\cite{David}. For chain length $N=32$ we anticipate rather large deviations from the SCF calculations at incompatibilities smaller than $\chi N=25$. Since for the present model the transition point 
is not known to high precision, we did not attempt a quantitative comparison of the results in Fig.\ref{fig:chic} with theoretical results for the order parameter profiles for surface--induced 
ordering\cite{SURFACE}.

At stronger incompatibilities $\epsilon_\Theta>0.5$, there
occurs a phase separation into a homopolymer-rich phase and a phase rich in vacancies. This would correspond to $\chi N \approx 85$ according to Eq.(\ref{eqn:chi}). However, we do
not expect this equation to hold in this temperature regime, because the fluid structure at these temperatures differs significantly from the high temperature structure. At much smaller incompatibilities, 
already, the width of the internal interfaces becomes comparable to the length scale of the local polymer architecture. Previous studies have shown that in the strongly segregated regime
the Gaussian chain model may yield qualitatively erroneous results\cite{MW}. These two considerations yield a rough estimate for the temperature interval in which good agreement between the MC simulations
and the SCF calculations can be expected.

We have quenched three different systems from their athermal state ($\epsilon=0$) to $\epsilon=0.1769k_BT$. The systems have the geometry $30 \times 96 \times 96$, $46 \times 93 \times 93$ and 
$56 \times 96 \times96$ in units of the lattice spacing. According to the SCF calculations these geometries correspond to the $L_2$, the $L_\perp$, and the $L_4$ phase, respectively. The lateral 
extension in the $L_\perp$ phase has been chosen compatible with the lamellar spacing in the SCF calculations. For each of these geometries we have simulated at least 4 independent systems. 
We do not find transitions from one morphology to a different one during the simulation time
which exceeds at least $3.8 \cdot 10^6$ Monte Carlo steps for each system. Hence, we cannot rule out metastability effects completely.  However, the 
simulation time is long enough that the observed structures are free of defects (on the scale of the simulation cell) and the composition profiles of 
systems which assembled into the same morphology agree (cf.\ also \cite{PAPER2}). 

\begin{figure}[htbp]
    \begin{minipage}[t]{160mm}%
       \mbox{
        \setlength{\epsfxsize}{7cm}
        \epsffile{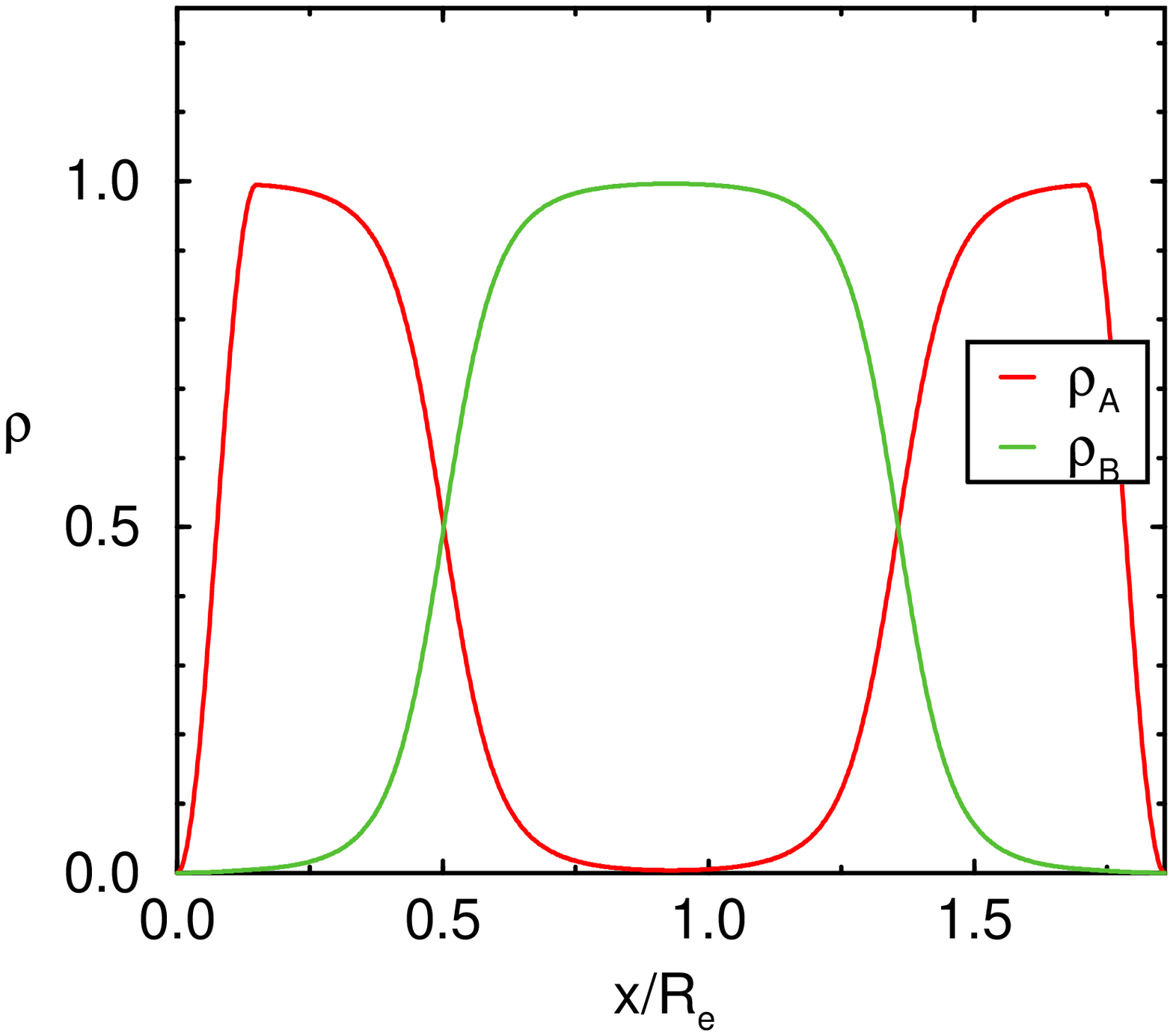}
        \setlength{\epsfxsize}{7cm}
        \epsffile{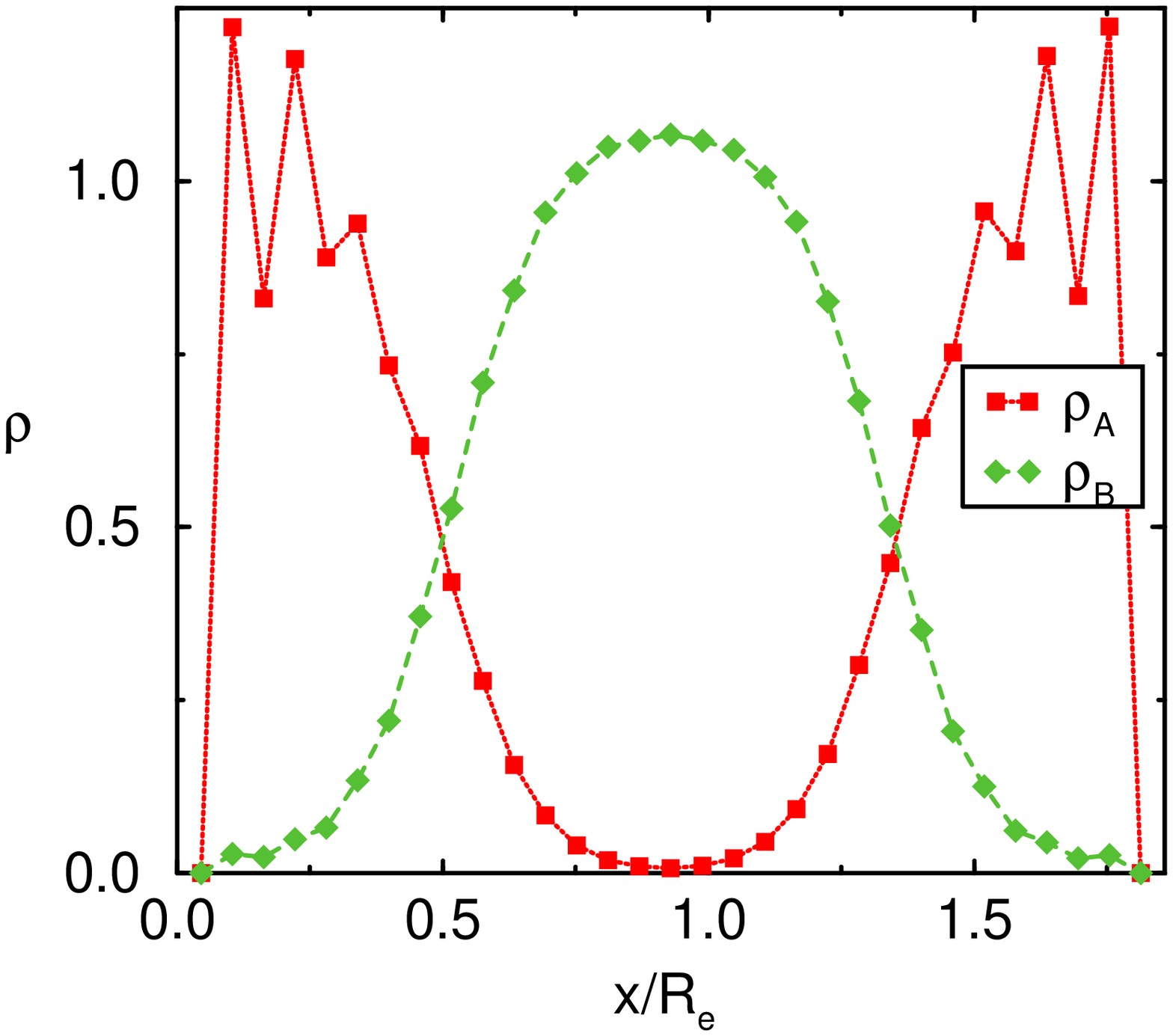}
       }
    \end{minipage}%
    \hfill%
    \begin{minipage}[b]{160mm}%
    \vspace*{1cm}
       \caption{
       \label{fig:raw} $A$ and $B$ density profiles of the $L_2$ phase in a symmetric film $\Delta/R_e=1.71$ and $\Lambda_1 N = \Lambda_2 N = 0.375$.
       ({\bf a}) results of the self-consistent field theory and
       ({\bf b}) results of the Monte Carlo simulations. Note the packing effects at the wall.
       }
    \end{minipage}%
\end{figure}

In Fig.\ref{fig:raw} we present the monomer density profiles of the $L_2$ phase in the SCF calculations ({\bf a}) and in the MC simulation ({\bf b}). 
Qualitatively, the profiles are similar. Both data sets show  almost completely segregated $A$ and $B$ rich regions separated via two interfaces. However, in the  SCF
calculations the total density rises smoothly from zero at the film boundaries to one in the middle of the film according to Eq.(\ref{eqn:dens}). In the MC simulations the monomers pack 
against the wall and produce oscillations in the density profile near the surfaces. These details of the
local fluid structure are not captured in the Gaussian chain model. Note that the average density in the two layers nearest to the walls is close to the bulk density. Thus the simple estimate for the
energy contribution of the walls (cf.\ Eq.(\ref{eqn:ewall})) is largely unaffected by the packing effects. 

Snapshots of the final morphologies are presented in 
Fig.\ref{fig:snap}. In panel ({\bf a}) we present the final snapshots for film thickness $\Delta_0=30$. The surfaces correspond to the top and bottom plane. All systems have assembled into the $L_2$ phase. In the snapshots 
the $A$ component corresponds to the darker species. Half an $A$ lamella is located at each wall. However, one also observes rather strong fluctuations which allow the $B$ component to protrude up to the wall.
All films of thickness $\Delta_0=46$ (cf.\ Fig.\ref{fig:snap}({\bf b})) have assembled into the $L_\perp$ phase. However, the simulations exhibit two different repeat distances. In the two systems, where the lamellae are oriented 
parallel to the 
box axis, the lamellae are spaced at a distance $D=1.827 R_e$, which agrees with the SCF calculations $D_{\rm SCF}=1.822 R_e$. However, in the six other systems, the lamellae make an angle with the box 
axis and the repeat distance is $D=1.938 R_e$. This value exceeds the SCF prediction by $6\%$. 
In the SCF calculations such a deviation from the prefered repeat distance increases the free energy by 0.01 $k_BT$ per molecule or 8 $k_BT$ for the whole system. Hence, the occurence of the
larger spacing cannot be explained by thermal fluctuations alone.  The difference might be traced back to the fact that we use for the SCF calculations the chain extension 
$R_e$ corresponding to the athermal state. As it has been observed in other simulations\cite{Fried} (cf.\ also Fig.\ref{fig:chic}), the chains stretch even in the disordered phase to avoid energetically unfavorable contacts between the different blocks. 
For the comparison with the SCF calculations\cite{PAPER2} we employ the two configurations in which the lamellae are oriented parallel to the box axis.
Three films of thickness $\Delta_0=56$ (cf.\ \ref{fig:snap}({\bf c})) assembled into the $L_4$ phase, whereas one system prefered the $L_\perp$ phase. For the comparison with the SCF calculations\cite{PAPER2} the latter system was discarded.
Though we cannot rule out that some of the systems are trapped in metastable conformations it is very gratifying that we observe the morphologies predicted by the SCF calculations except for one case 
at the largest film thickness.

\section{ Summary. }
We have presented SCF calculations and MC simulations for symmetric diblock copolymers confined into a thin film. Both surfaces attract the same component of the diblock via a short range
potential. We have calculated the phase diagram as a function of the incompatibility $\chi N$ and the film thickness in mean field approximation and discussed the stability
of parallel $L_p$ and perpendicular $L_\perp$ aligned lamellar phases. At high incompatibility we find the sequence $L_2$,$L_\perp$,$L_4$,$L_\perp$, $L_6$ while we find a direct transition between 
the $L_4$ and the $L_6$ phase at weak segregation.

At low incompatibilities we find rather pronounced deviations from the SCF theory. Most notably, for chain length $N=32$ the onset of ordering occurs around $13.5 \leq \chi_c N \leq 17$ instead of
$\chi_c^{\rm SCF} N \approx 10.5$. The order of magnitude of the shift is compatible with corrections to the mean field behavior\cite{Fredrickson,David}. At very high segregation, we expect the
local structure of the model to be important.

At $\chi N=30$ we have compared the results of the SCF calculations with MC simulations in the framework of the BFM. We find qualitative agreement between the MC simulations and the SCF calculations.
In particular, we observed the $L_2$, $L_\perp$ and $L_4$ phases as predicted by the SCF calculations. The main difference between the SCF calculations and the MC simulations is the structure
at the surfaces. While the density decays smoothly to zero in the SCF calculations, there are pronounced packing effects in the MC simulations. In both schemes the incompatibility at the film surface is
reduced. In the SCF calculations it stems from the reduced density at the surface, while in the MC simulations it is due to the finite extension of the interactions (``missing neighbor effect'').
Both effects give rise to a negative line tension\cite{MSCF} as the interface approaches the surface and lead to a stabilization of the $L_\perp$ phase. The effect is, however, more 
pronounced in the MC simulations. Another difference between the MC simulations and the SCF calculations is the dependence of the chain extension on the incompatibility.
The majority of systems in the $L_\perp$ phase has assembled into a morphology with a lamellar spacing which exceed the prediction of the SCF calculations by about $6\%$. This goes along with
a stretching of the  diblock copolymers already above the ordering transition. The effect has been observed in previous simulations\cite{Fried,Sommer} and experiments\cite{Maurer,Stamm1,Stamm}, and can be 
rationalized via a coupling of the intramolecular energy and the chain conformations. 

In view of these effects the agreement with the SCF calculations is satisfactory.
A detailed comparison of individual profiles between the SCF calculations and the MC simulations shall be presented in the following paper\cite{PAPER2}.

\subsection*{Acknowledgment}
It is a great pleasure to thank P.K. Janert for discussions and technical advice. We have also benefited from stimulating discussions/correspondence with F. Schmid
and M.W. Matsen. We acknowledge generous access to the CRAY T3E at the HLR Stuttgart and HLRZ J{\"u}lich, as well as access to the CONVEX SPP at the computing center 
in Mainz.  Financial support was provided by the DFG under grant Bi314/17.

\begin{figure}[htbp]
    \begin{minipage}[b]{160mm}%
    \vspace*{1cm}
       \caption{
       \label{fig:snap} 
       Snapshot of the morphologies found in the MC simulations by quenching the systems from $\chi N=0$ to $\chi N=30$.
       ({\bf a}) $\Delta_0=30$
       ({\bf b}) $\Delta_0=56$
       ({\bf c}) $\Delta_0=46$
       Figures available upon request.
       }
    \end{minipage}%
\end{figure}


\begin{thebibliography}{99}
\bibitem{Leibler} L. Leibler, Macromolecules {\bf 13}, 1602 (1980).
\bibitem{Semenov} A.N. Semenov, Soviet Physics JETP {\bf 61}, 733 (1985).
\bibitem{Matsen}  M.W. Matsen and M. Schick, Phys.Rev.Lett. {\bf 74}, 4225 (1995).
\bibitem{TODAY} F.S. Bates and G.H. Fredrickson, Physics Today {\bf 52}, 33 (1999).
\bibitem{SCF1}    E. Helfand and Y. Tagami,  J.Chem.Phys. {\bf 56}, 3592 (1972).
\bibitem{SCF2}    J. Noolandi and K.M. Hong, Macromolecules {\bf 14}, 727 (1981).
\bibitem{SCF3} for a recent review see e.g., F.Schmid, J.Phys.:Condens.Matt. {\bf 10}, 8105 (1998).
\bibitem{Russell} T.P Russell, Current Opinion in Colloid \& Interfacial Science {\bf 1}, 107 (1996).
\bibitem{Lambooy} P. Lambooy, T.P. Russell, G.J. Kellog, A.M. Mayes, P.D. Gallagher, and S.K. Satija, Phys.Rev.Lett. {\bf 72}, 2899 (1994)
\bibitem{Kellogg} G.J. Kellogg, D.G. Walton, A.M. Mayes, P.Lambooy, T.P. Russell, P.D. Gallagher, and S.K. Satija, Phys.Rev.Lett. {\bf 76}, 2503 (1996).
\bibitem{Walton} D.G. Walton, G.J. Kellogg, A.M. Mayes, P. Lambooy, and T.P Russell, Macromolecules {\bf 27}, 6225 (1994).
\bibitem{Pickett} G.T. Pickett and A.C. Balazs, Macromolecules {\bf 30}, 3097 (1997).
\bibitem{MSCF} M.W. Matsen, J.Chem.Phys. {\bf 106}, 7781 (1997).   
\bibitem{REVIEW1}  M.W. Matsen Current Opinion in Colloid and Interfacial Science {\bf 3}, 40 (1998).
\bibitem{Kikuchi} M. Kikuchi and K. Binder, J.Chem.Phys. {\bf 101}, 3367 (1994); Europhys.Lett. {\bf 21}, 427 (1993).
\bibitem{Koneripalli} M. Koneripalli, R. Levicky, F.S.Bates, J. Ankner, H. Kaiser, and S.K. Satija, Langmuir {\bf 26}, 6681 (1996).
\bibitem{REVIEW2} K. Binder, Adv.Pol.Sci {\bf 138},1 (1999).
\bibitem{BFM} I. Carmesin and K. Kremer, J.Phys. (France) {\bf 51}, 915 (1990); H.-P. Deutsch and K. Binder, J.Chem.Phys. {\bf 94}, 2294 (1991).
\bibitem{BATESPRL}  N. Koneripalli, F.S. Bates and G.H. Fredrickson, Phys.Rev.Lett. {\bf 81}, 1861 (1998).
\bibitem{C1}    There is a typographical mistake in the corresponding equation in \cite{MSCF}.
\bibitem{MCONF} M. M{\"u}ller, Macromolecules {\bf 31}, 9044 (1998).
\bibitem{Fried} H. Fried and K. Binder, Euro.Phys.Lett. {\bf 16}, 237 (1991); J.Chem.Phys. {\bf 94}, 8349 (1991); Macromolecules {\bf 26}, 6878 (1993).
\bibitem{Maurer} W.W. Maurer, F.S. Bates, T.P. Lodge, K. Almdal, K. Mortensen, and G.H. Fredrickson, J.Chem.Phys. {\bf 108}, 2989 (1998).
\bibitem{Stamm1} K. Almdal, J.H. Rosedale, F.S. Bates, G.D. Wignall, and G.H. Fredrickson, Phys.Rev.Lett. {\bf 65}, 1112 (1990);\\
                 J.H. Rosedale, F.S. Bates, K. Almdal, K. Mortensen, and G.D. Wignall, Macromolecules {\bf 28}, 1429 (1995).
\bibitem{Stamm}  V.T. Bartels, M. Stamm, V. Abetz, and K. Mortensen, Euro.Phys.Lett. {\bf 31}, 81 (1995).
\bibitem{MREV} for a review see M. M{\"u}ller, Macromolecular Theory and Simulation (in press).
\bibitem{OLD} F. Schmid and M. M{\"u}ller, Macromolecules {\bf 28}, 8639 (1995);\\
              M. M{\"u}ller, K. Binder, and W. Oed, J.Chem.Soc. Faraday Trans. 91, 2369 (1995).
\bibitem{MW}     M. M\"uller and A. Werner, J.Chem.Phys. {\bf 107}, 10764 (1997).
\bibitem{REV}  M. M\"uller and F. Schmid, in {\em Annual Reviews of Computational Physics},  D. Stauffer (ed.) in press.\\
		     (also available from http://xxx.lanl.gov/abs/cond-mat/9803111).
\bibitem{MS1}   M. M\"uller and M. Schick, J.Chem.Phys. {\bf 105}, 8885 (1996).
\bibitem{M0}  M. M{\"u}ller and K. Binder, Macromolecules {\bf 28}, 1825 (1995).
\bibitem{WET} M. M\"uller and K. Binder, Macromolecules {\bf 31}, 8323 (1998).
\bibitem{Grest} G.S. Grest, M.-D. Lacasse, K. Kremer, and A. Gupta, J.Chem.Phys. {\bf 106}, 6709 (1997).
\bibitem{Turner} M.S. Turner, Phys.Rev.Lett. {\bf 72}, 2899 (1994).
\bibitem{MWC} S.T. Milner, T.A. Witten, and M.E. Cates, Macromolecules {\bf 21}, 2160 (1988).
\bibitem{Lifshitz} I.M. Lifshitz, A.Y. Grosberg, and A.R. Kohklov, Rev.Mod.Phys. {\bf 50}, 683 (1978).
\bibitem{Williams} G.G. Pereira and D.R.M. Williams, Macromolecules {\bf 32}, 1661 (1999).
\bibitem{Chaikin} M. Park, C. Harrison, P.M. Chaikin, R.A. Registar, and D.H. Adamson, Science {\bf 276}, 1401 (1997);\\
    P. Mansky, C. Harrison, P.M. Chaikin, R.A. Registar, Applied Physics Letters {\bf 68}, 2586 (1996).
\bibitem{Fredrickson} G.H. Fredrickson and E. Helfand, J.Chem.Phys. {\bf 87}, 697 (1987).
\bibitem{C2}    A precise estimate of the transition temperature is difficult, because the surface suppresses fluctuations and 
		this might lead to a systematic overestimation of the transition temperature of the bulk system.
\bibitem{David}  E.F. David and K.S. Schweizer, J.Chem.Soc.Faraday Trans. {\bf 91}, 2411 (1995).
\bibitem{SURFACE} G.H. Fredrickson, Macromolecules {\bf 20}, 2535 (1987);
		  H. Tang and K.F. Fried, J.Chem.Phys. {\bf 97}, 4496 (1992);
		  K. Binder, H.L. Frisch, and S. Stepanow, J.Phys.II (France) {\bf 7}, 1353 (1997).
\bibitem{Sommer}  A. Hoffmann, J.U. Sommer, and A. Blumen, J.Chem.Phys. {\bf 107}, 7559 (1997).
\bibitem{PAPER2} T. Geisinger, M. M\"uller, and K. Binder, following paper.


\end{thebibliography}
\end{document}